# Experimental proof for resonant diffusion of normal alkanes in LTL and ZSM-12 zeolites

K. Yoo, R. Tsekov and P.G. Smirniotis

Department of Chemical and Material Engineering, University of Cincinnati, Cincinnati, USA

The intra-crystalline diffusion of normal alkanes in LTL and ZSM-12 zeolite was experimentally studied via gravimetric measurements performed at different temperatures. A periodic dependence of the diffusion coefficient on the number of carbon atoms in alkane was detected, which is an experimental proof for resonant diffusion. The present observations were described on the base of the existing theory of the resonant diffusion and several important parameters of the alkane-zeolite interaction and zeolite vibrations were obtained. In the considered temperature region the diffusion coefficient follows the Arrhenius law with periodic dependences of the pre-exponential factor and activation energy on the number of carbon atoms in alkanes. A compensation effect of simultaneous increases of the pre-exponential factor and the activation energy was also established.

Diffusion of hydrocarbons in zeolites has been extensively studied both with experimental and theoretical approaches. This is due to the large number of applications of these materials in the refinery industry, e.g. as catalysts in hydroisomerization and cracking processes [1] and as molecular sieves in separation processes, such as pressure swing adsorption [2] and membrane-based separation [3]. A significant factor for the efficiency of these processes is the rate of intracrystalline diffusion of species in zeolites. In all cases, the dynamic behavior of sorbed molecules is crucial in determining the performance of the considered catalytic or separation process. An interesting feature of the diffusion of normal alkanes through zeolites is the resonant diffusion, when the diffusion coefficient exhibits a periodic dependence on the number of carbon atoms of the alkane chain. This behavior was first reported by Gorring [4] who measured the macroscopic diffusivity for a series of normal alkanes in zeolite T. Other researchers [5], however, were unsuccessful to observe experimentally this sort of periodicity of the diffusion constant. This apparent discrepancy could be explained by the theory of resonant diffusion [6, 7], which elucidates the necessary conditions for the zeolite structure and alkane-zeolite interaction in order to observe a periodic dependence of the diffusion coefficient on the number of carbon atoms in alkanes.

In our previous paper [8], the diffusion of normal alkanes in one-dimensional zeolites is theoretically studied using the stochastic equation formalism. The calculated diffusion coefficient accounts for the vibrations of the diffusing molecule and zeolite framework, alkane-zeolite interaction, and specific zeolite structure. It is shown that if the interaction potential is modulated by the zeolite nano-pore structure, the diffusion coefficient varies periodically with the number of carbon atoms of the alkane molecule. The present paper is an experimental proof of this theory and one of the seldom examples when the resonant diffusion holds.

*Material Preparation*

LTL possessing a one-dimensional pore with aperture (of 7.1 Å) connected with large void spaces (of 12.6 Å) was kindly provided by UOP. ZSM-12, possessing a one-dimensional non-interconnecting tubular-like channel structure (pore size of 5.6 × 6.1 Å) [9] was synthesized hydrothermally following the procedure described in our earlier studies [10]. The synthesized zeolites were calcined in air at 520 °C for 4 hours to remove the occluded template. In order to minimize the hydrophilic nature of zeolite sample, the samples were prepared as Na-form zeolite. The Na-form of each zeolite was obtained by performing a cation exchange of the charge balancing alkali metal with $Na^+$. This procedure involved heating the sample in a 2.0 M NaCl solution at 90 °C for 10 hours under reflux.

*Material Characterization*

The structural properties of the samples were obtained from the nitrogen isotherms at 77 K measured in a Micromeritics ASAP 2010 apparatus. Before measurements, all samples were degassed with He for 2 hours at 200 °C. Scanning electron microscopy was done on selected samples to determine the crystallite size and morphology using a Hitachi S-4000 Field Emission SEM. The size of zeolite crystallites was determined with a laser scattering particle size analyzer (Malvern Mastersizer S series). For this experiment, the wet method was used as the medium for dispersion of the zeolite. The solution was ultrasonicated for 30 min in order to break down the flocculates before the run was carried out. All samples were run twice to ensure the accuracy of the measurements. The bulk Si/Al ratios of the zeolites were determined using a Thermo Jarrell Ash Inductively Coupled Plasma–Atomic Emission spectrometer (ICP-AES).

*Procedure*

The gravimetric measurements were carried out using by modified thermo-gravimetric analysis (Perkin-Elmer TGA7) system with continuous monitoring of the mass changes. A small amount of each sample (about 10 mg), hence a very small thickness of the zeolite layer was loaded on the microbalance, which enabled us to avoid non-isothermictity. The sample was activated for 1 h at 550 °C and then cooled down to desirable temperature. After the change of the weight over sample was constant, each hydrocarbon was pumped with carry gas (He, purity <99.9%) controlled by mass flower controller. The microbalance weight measurements were accurate to about $10^{-5}$ mg, with data acquisition software (Pyris).

For an isothermal system the solution for the transient diffusion equation for a spherical particle with radius $R$ is [11]:

$$\Delta m_t = \Delta m_\infty [1 - 6/\pi^2 \sum_{j=1}^{\infty} \exp(-j^2 \pi^2 Dt/R^2)/j^2] \tag{1}$$

where $\Delta m_t$ is the mass increase for time $t$ and $D$ is the diffusion constant. For long times, this solution reduces to the following approximate form [12]:

$$\Delta m_t = \Delta m_\infty [1 - 6/\pi^2 \exp(-\pi^2 Dt/R^2)] \tag{2}$$

Thus, eq (2) predicts a linear relationship between $\ln(1 - \Delta m_t / \Delta m_\infty)$ and time in the long time region. The slope of this straight line yields directly the intra-crystalline diffusion coefficient $D$.

*Material Characterization*

The structural properties of the samples used in this study are given in Table 1. Both samples have similar structural properties such as BET area, pore volume, and mean diameter of particle. A scanning electron micrograph of LTL and ZSM-12 obtained is shown in Fig. 1. The crystal size of each zeolite was comparable with the value obtained by a laser scattering particle size analyzer.

**Table 1.** Structural properties of zeolite samples

| Zeolite | Si/Al (Bulk)[a] | BET area (m²/g) | Pore Volume (cm³/g) | Mean diameter (μm) |
|---|---|---|---|---|
| LTL | 3.5 | 306 | 0.23 | 0.85 |
| ZSM-12 | 31 | 320 | 0.25 | 1.10 |

[a] Determined by chemical analysis

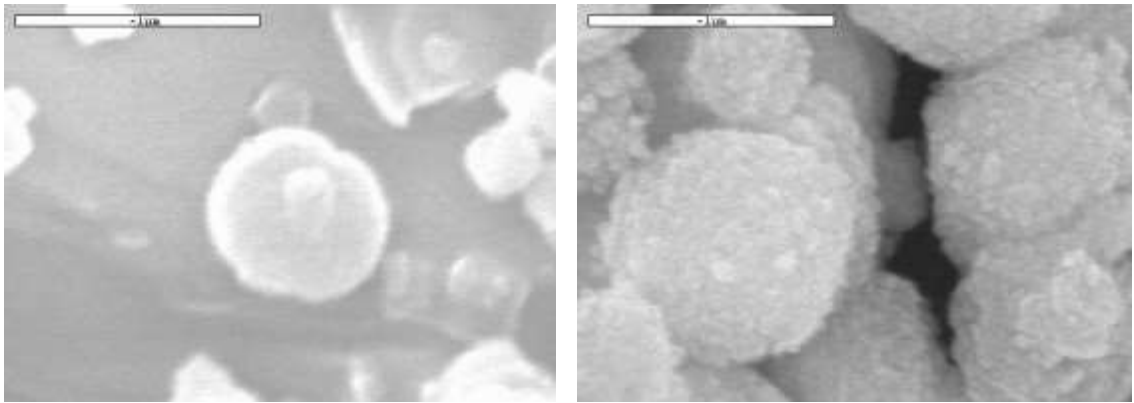

**Fig. 1**. Scanning electron micrograph of (a) LTL and (b) ZSM-12

*Theory*

The diffusion of normal alkanes through one-dimensional zeolite networks was theoretically described in a previous paper [8]. As a result the following expression for the diffusion coefficient $D$ of alkanes was derived

$$D = 1/[\int_0^1 \beta B \exp(\beta U) d\varphi \int_0^1 \exp(-\beta U) d\varphi] \tag{3}$$

where $1/\beta = kT$ is the thermal energy. The friction coefficient $B$ and the effective interaction potential of the alkane molecule with the zeolite $U$ are given by

$$B = 24\pi^5 C^2 N \cos^2(2\pi n\varphi) / \mu \omega_0^3 a^4 \tag{4}$$

$$U = A\cos(2\pi\varphi)\sin(\pi N/n)/\sin(\pi/n) + CN\cos(2\pi n\varphi) - 6\pi^2 C^2 N \sin^2(2\pi n\varphi)/\mu \omega_\infty^2 a^2 \tag{5}$$

where $N$ is the number of carbon atoms in the alkane molecule, $A$ and $C$ are the half-height of the energy barriers due to the channel and atomic structures of the zeolite, respectively, $\mu = 20$ g/mol is the average mass of the zeolite atoms, $a = 1.25$ Å and $na$ are the period lengths of the atomic and channel structures, respectively, and $\omega_0$ and $\omega_\infty$ are specific frequencies of the zeolite vibrations defined via

$$\omega_0^{-3} = f(0) \qquad \omega_\infty^{-2} = \int_0^\infty f(\omega) d\omega \tag{6}$$

where $3\omega^2 f(\omega)$ is the phonon density in the zeolite. The first term in eq (5) depends periodically on the number of carbon atoms of the alkane. It is zero if the ratio *N/n* is an integer number and its absolute value is maximal if the ratio *N/n* is a half number. This is a precondition for resonant diffusion. The last term in eq. (5) is important only if the inequality $\mu \omega_\infty^2 a^2 \leq 6\pi^2 C$ holds. Since the frequency $\omega_\infty$ is of the order of the Debye frequency of the zeolite, this inequality is usually not satisfied. Hence, in the further consideration the last term in eq (5) will be neglected.

Since the friction coefficient $B$ according to eq (4) depends linearly on $N$, a reasonable way to detect the resonant parameters is to analyze the product $ND$ as a function of $N$. Doing so, one can observe from the present experimental data a strong maximum at $N = 7$ for either LTL or ZMS-12 zeolites. This means that in both cases $n$ should be equal to 7. In the case of LTL, however, the ratio between the periods of the channel and atomic structures is about 6. This slight discrepancy could be due to the conformation changes of the alkane molecule during the diffusion process. Adopting these two numbers, reasonable approximations for the friction coefficient and potential energy of alkanes in LTL are

$$B = 24\pi^5 C^2 N \cos^2(12\pi\varphi) / \mu\omega_0^3 a^4 \tag{7a}$$

$$U = A\cos(2\pi\varphi)\sin(\pi N/7)/\sin(\pi/7) + CN\cos(12\pi\varphi) \tag{7b}$$

The situation in ZSM-12 is a little bit complicated. From the data of the zeolite structure it follows that $n$ should be equal to 4. Since this is much smaller than 7, our conclusion is that in ZMS-12 the effective channel potential has a period twice larger than $4a$. This means that the diffusion path of the alkane molecule feels every second ring, which could be due to the non-symmetric distribution of the atoms in the rings. Hence, the friction coefficient and the effective potential in ZSM-12 can be approximated by

$$B = 24\pi^5 C^2 N \cos^2(16\pi\varphi) / \mu\omega_0^3 a^4 \tag{8a}$$

$$U = A\cos(2\pi\varphi)\sin(\pi N/7)/\sin(\pi/7) + CN\cos(16\pi\varphi) \tag{8b}$$

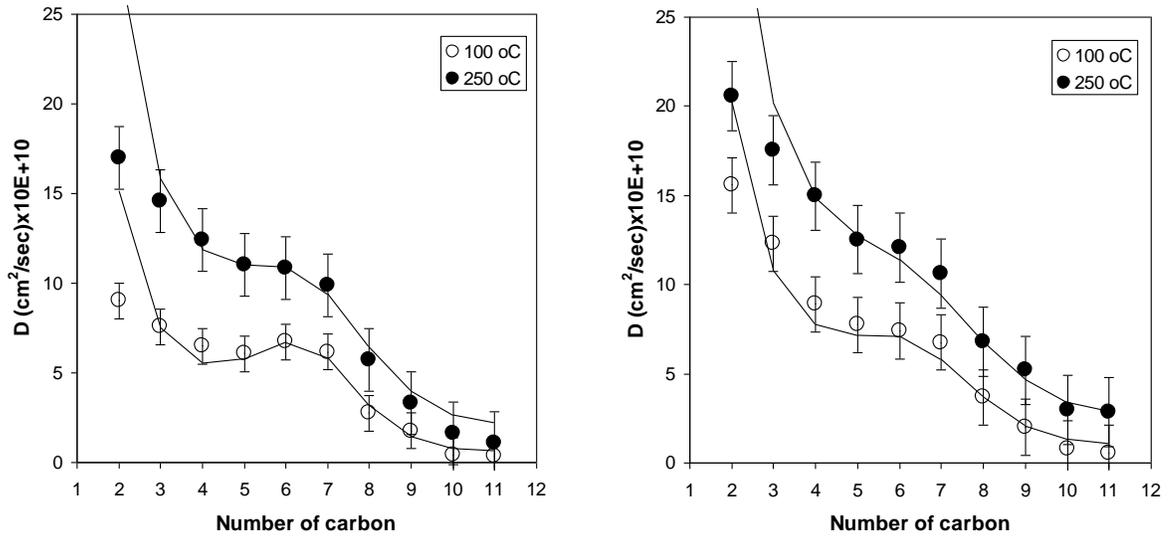

**Figs. 2-3**. Dependence of the diffusion coefficient of normal alkanes in LTL and ZSM-12 zeolite on the number of carbon atoms at two different temperatures

The parameters $\omega_0$ and $C$ could be considered as common for the both zeolites. The parameter $A$, however, is different since the structure of the rings in LTL and ZSM-12 is not the same. Substituting eq (7) or eq (8) in eq (3) one can calculate the diffusion coefficients of alkanes in LTL or ZSM-12, respectively. In Figs. 2 and 3 the best fits of the experimental results are presented which corresponds to the following values of the variable parameters: $\omega_0 = 230$ GHz, $C = 0.3$ kJ/mol, $A = 1.8$ kJ/mol (LTL) and $A = 1.6$ kJ/mol (ZSM-12). As one can observe, the theory reproduces well the experimental data. The only exception is the ethane. Because the ethane molecule is very small it can rotate in the zeolite and in this way to violate the resonant condition. As a result the experimentally measured diffusion coefficient of ethane is much lower than the theoretical prediction. In any case the energy $C$ is an order of magnitude lower than $A$ which is due to the fact that the distance between the zeolite atoms and the alkane molecule in the openings is larger than that in the rings. The value of $\omega_0$ is also quite reasonable in comparison to the typical values of the Debye frequencies in zeolites.

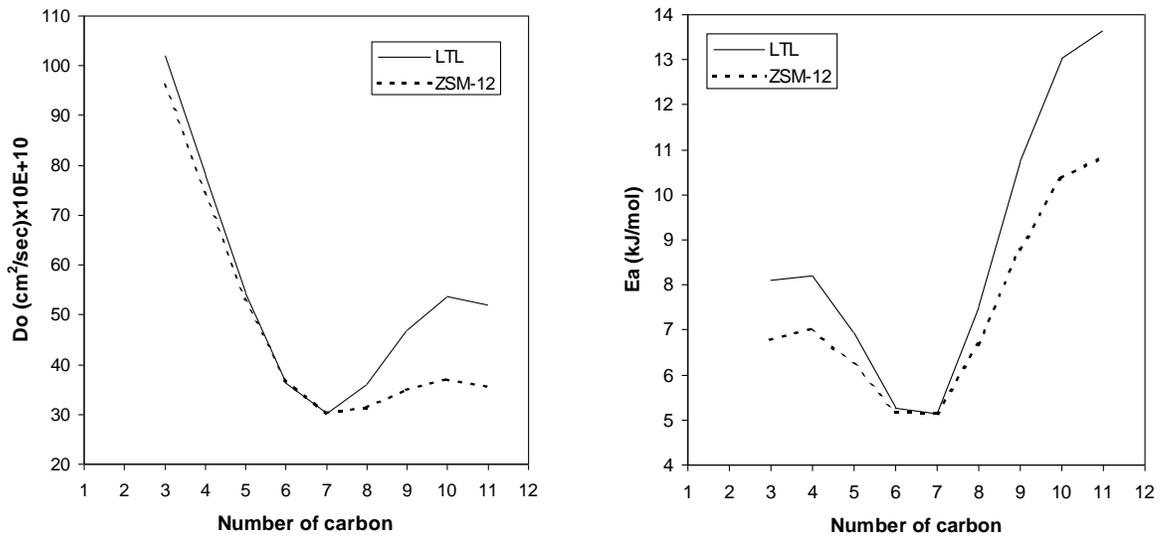

**Figs. 4-5.** Dependence of the pre-exponential factor and the activation energy of the diffusion coefficient of normal alkanes in LTL and ZSM-12 zeolite on the number of carbon atoms

A further study of the dependence of the diffusion coefficient on temperature unveils that for both LTL and ZSM-12 zeolites in the considered temperature interval $D$ obeys the Arrhenius law

$$D = D_0 \exp(-\beta E_a) \tag{9}$$

The corresponding resonant dependences of the pre-exponential factor $D_0$ and activation energy $E_a$ on the number $N$ of carbon atoms are presented in Figs. 4 and 5. As seen both the pre-exponential factor and activation energy exhibit minima at $N = 7$ due to the resonant depend-ence of the interaction potential. It is important to note that there is a compensation effect corresponding of increase of $D_0$ when $E_a$ increases. In such a way the unfavorable effect of higher activation energy is partially reduced by a larger pre-exponential factor.

The present experimental observation of a non-monotonous dependence of the diffusion coefficient of normal alkanes in LTL and ZSM-12 zeolites on the number of carbon atoms in the alkane molecule is an undoubted proof for resonant diffusion, since it is explained very well by the relevant theory. Hence, the resonant diffusion should not be further ignored in the science and practice. It could play an important role in the catalytic kinetics and especially in the separation processes, where the diffusivity of the species is of crucial importance. In general the resonant effects on diffusivity should be present in all structured materials. The problem is that usually other dominant factors prohibit the detection of resonant diffusion. Due to the very simple structure of the LTL and ZSM-12 one-dimensional channels the present test succeeds to monitor the resonant effects.